**Analyzing a Complex Game for the South China Sea Fishing Dispute using Response Surface Methodologies**

Michael Perry, mperry20@gmu.edu
George Mason University, Department of Systems Engineering and Operations Research, Fairfax, VA, USA
ORCID: 0000-0001-7911-4083

**Abstract**
The South China Sea (SCS) is one of the most economically valuable natural resources on the planet, and as such has become a source of territorial disputes between its bordering nations. Among other things, states compete to harvest the multitude of fish species in the SCS. In an effort to gain a competitive advantage states have turned to increased maritime patrols, as well as the use of "maritime militias," which are fishermen armed with martial assets to resist the influence of patrols. This conflict suggests a game of strategic resource allocation where states allocate patrols intelligently to earn the greatest possible utility. The game, however, is quite computationally challenging when considering its size (there are several distinct fisheries in the SCS), the nonlinear nature of biomass growth, and the influence of patrol allocations on costs imposed on fishermen. Further, uncertainty in player behavior attributed to modeling error requires a robust analysis to fully capture the dispute's dynamics. To model such a complex scenario, this paper employs a response surface methodology to assess optimal patrolling strategies and their impact on realized utilities. The methodology developed successfully finds strategies which are more robust to behavioral uncertainty than a more straight-forward method.

# 1 Introduction

The South China Sea (SCS) is one of the most profitable bodies of water for commercial fishing and has become the source of territorial disputes between bordering countries. Traditional United Nations defined exclusive economic zones (EEZs), which extend 200 nautical miles beyond a state's coastline, are problematic in a congested maritime environment such as the SCS where China, Vietnam, Malaysia, Brunei, the Philippines, and Taiwan have all made territorial claims which overlap with those of at least one other state. This situation has led to more than mere diplomatic jostling; real, physical clashes have occurred between fishermen and the coast guards and navies of rival nations, and in the most extreme cases have led to the loss of life. The economic implications are also enormous as encroachments by one nation's fishermen into waters claimed by another are a daily occurrence. The parties in question have taken actions to improve their standing in the dispute. On the one hand, some countries have equipped their fishermen with martial assets as a form of "maritime militia" to mitigate the deterrent effect of rival maritime patrols. On the other, all nations other than China appear willing to cooperate to resolve the dispute via a multilateral fisheries management organization called the Southeast Asian Fisheries Development Center (SEAFDEC). The U.S. has also taken an interest in the dispute on account of its rivalry with China and the broader implications for control of the SCS.

Given the SCS is vast and composed of multiple fisheries, it's natural to model this dispute as a many-variate game of strategic resource allocation. Each fishery can be modeled using a bionomic model and the additional costs imposed by maritime patrols can be informed by recent empirical research on the subject. There are multiple modeling challenges, however. First, even the simplest bionomic models won't lead to analytic solutions when multiple fisheries are in dispute, so sampling methods will be required. Second, decently sized problems become computationally intensive and thus efficiency of sampling will be paramount. While not commonly used in the game theory literature, this first pair of problems can be addressed using well-established techniques in response surface methodologies (RSM). A third problem is a bit more nuanced and requires ingenuity.

Traditionally, a game-theoretic model would predict the players' behavior using a Nash equilibrium, or near-Nash equilibrium, but this can hide key risks when deviations from such model-prescribed behavior leads to large deviations in realized utilities. Deviations in behavior could result for several reasons:

1. Foremost, the underlying model is almost surely not a perfect representation of the player's utility functions, and deviations should thus be expected.
2. Even if the model were perfect, if the players being modeled don't possess the tools to find true optimal responses they'll rather "satisfice" by finding a good solution that might be suboptimal. This creates a host of possible strategies one player may employ in response to the strategies of all others, rather than a single optimal response and in turn a single pure Nash equilibrium.

This paper introduces a novel formulation allowing for reasonable deviations from model-defined optimal responses, referred to as "behavioral uncertainty" for the remainder of the paper. As will be explicated, this formulation is akin to near-Nash equilibria. The distinctive feature of the formulation presented here is that it's robust, allowing a decision maker to strategize against worst-case scenarios regarding an adversary's deviation from his model-defined optimal response. This added layer of complexity heightens the need for efficient sample selection. Also distinct from near-Nash equilibria, the model formulation assumes sequential play. The relative merit of this assumption as a decision-making tool is briefly discussed in Section 4.1, but is not the focus of this paper. The focus is the computational challenge introduced by the robust formulation to address behavioral uncertainty and the complexity and size of the underlying fisheries model.

The remainder of the paper is organized as follows. Section 2 reviews the literature on the SCS dispute, fishery games, concepts related to behavioral uncertainty, and response surface methodologies in general. Section 3 presents the model for the optimal response of a player allocating patrols across fisheries, which is assumed to be arbitrarily complex and solvable only via a derivative-free, computationally expensive optimizer. Section 4 describes a robust game a decision-maker may use to analyze her patrol allocation problem and a sampling algorithm to estimate it's solution. Section 5 provides examples to show the algorithm's performance, and Section 6 concludes and comments on future research.

## 2 Literature Review

The South China Sea accounts for approximately one-tenth of all fish caught globally, making it one of the most economically important natural resources on the planet ("South China Sea Threatened by 'a Series of Catastrophes'" 2019). More broadly, the SCS is economically and politically important due to its strategic implications for shipping, energy resources, and its potential military importance during war (Buszynski 2012; Yoshihara 2012). Even under a more mild set of conditions, geography alone would likely lead to disputed claims of ownership as traditionally defined EEZs overlap.  In fact, each China, Vietnam, Malaysia, Brunei, the Philippines, and Taiwan have made claims that overlap with those of at least one other nation (Stearns 2012).  The competitive environment in the SCS has contributed to the rapid growth in China's coast guard to the point where it's larger than that of its neighbors combined (Erickson 2018), and to the expansion of state-sponsored maritime militias, which are commercial fishermen armed with small-arms and water cannons intended to offset the deterrent effects of patrols (Erickson and Kennedy 2016; Zhang and Bateman 2017).  Instances of fishermen clashing with coast guard patrols at sea are common; often, rival coast guards will confront one another in response to a distress signal sent by a fisherman (China Power Team 2020).  While this is a multilateral dispute, it's often thought of through a China-versus-the-rest lens.  This isn't entirely inappropriate, especially when viewed in terms of the implications for fisheries management and the emergence of SEAFDEC, an intergovernmental fisheries management organization including the major SCS nations other than China.  This paper will take this view and present a two-player game.

From a modeling perceptive, bionomic models of fisheries represent a well-established field and these models have been applied to relatively simple games to draw general insights. Introductory texts on bionomic fisheries models and game theory in fisheries management are (Clark 2006) and (Grønbæk et al. 2020), respectively.  (Grønbæk et al. 2020) covers noncooperative and cooperative games, two- and many-player games, asymmetries, and discounted payoffs.  Advanced complicating factors such as multiple interacting species (Fischer and Mirman 1996), coalition formation in many-player games (Long and Flaaten 2011), and uncertainty in stock levels and growth rates (Miller and Nkuiya 2016) have also been analyzed using games. The most general conclusion drawn is that, as in other games modeling common-pool resources, fish become overexploited in a competitive environment.

Optimal maritime patrol allocation has also been analyzed using game theory, though not through the lens of territorial disputes. Rather, these models have taken fishing rights as given and determined how to allocate a single state's patrols to combat illegal fishing (Fang, Stone, and Tambe 2015; Brown, Haskell, and Tambe 2014). A related strand of research has conducted empirical analysis on the extent to which patrols (and other factors) can deter illegal fishermen (Petrossian 2015). Crucially, empirical analysis has deliberately left out data on the SCS because it's not clear the same relationships will govern undisputed and disputed EEZs. Consider, for instance, the ability of maritime militias and interacting patrols to offset the effects of a single state's patrols. The true nature of the effect of patrols in the SCS is therefore an open question; this paper presents a simple model that in theory could be estimated with data, but that empirical research is left for follow-on research.

What was defined as behavioral uncertainty in Section 1 has been studied in a variety of ways. For instance, quantal response models assume players are more likely to play strategies with higher utilities, and don't require that they play the true optimal strategy with probability 1 (McKelvey and Palfrey 1995; Nguyen et al. 2013). In recent years adversarial risk analysis (ARA) has emerged as a method incorporating uncertainty in the model parameters to derive probabilistic distributions of players' behaviors; this derivation generally occurs by drawing Monte Carlo samples for the model parameters and then fitting an empirical distribution to realized behavior (Rios Insua, Rios, and Banks 2009; Banks, Rios, and Rios Insua 2016). The approach used in this paper to address behavioral uncertainty is both distribution- and parametric-free, making it akin to robust analysis. Robust games, generally defined as assessing the worst-case scenario for an adversary's behavior, have been addressed from a theoretical perspective in (Aghassi and Bertsimas 2006), (Kardes 2005), and (Crespi, Radi, and Rocca 2020). Robustness has been applied to realistic, though small, games in (Brown, Haskell, and Tambe 2014) and (McLay, Rothschild, and Guikema 2012).

All the methodologies mentioned above to account for behavioral uncertainty either impose restrictive parametric forms on the game, do not scale well to large problems, or both, and hence the methodological focus of this paper on developing a response surface methodology to analyze large, complex games. RSM approximates a function that's time consuming to compute, such as a player's optimal response, using an instantaneously computable function which is calibrated to a small sample of realizations of the original function. RSM has long been popular for approximating difficult optimizations and a taxonomy of methods was presented by (Jones 2001). This taxonomy decomposes RSM into interpolating and non-interpolating methods, and one-stage and two-stage methods where the latter generates subsequent sample points following an initial sampling of the decision space. Polynomial regressions and other parametric models have traditionally been used as response surfaces but black-box models such as boosted regression trees and artificial neural networks can be incorporated as well (Hastie, Tibshirani, and Friedman 2009). Two-stage methods involve optimizing the response surface to pick subsequent sample points; this is feasible when using the aforementioned black-box models given cutting-edge global optimizers and increased computing power (Xu et al. 2015).

# 3 The Fishing Dispute Game with Multiple Fisheries

Two players, Blue and Red, are competing over $k$ fisheries and must allocate a fixed amount of maritime patrols to each. This is akin to the SCS where the largest state, China, is largely in competition with a coalition of the others. Section 3.1 introduces the fisheries model governing utility extracted from a fishery, and Section 3.2 states the optimal response problem of one player given the maritime patrol allocation of the other.

## 3.1 A fisheries model with costs imposed by patrols

The biomass, $x_i$, of each fishery $i = 1:k$ is modeled by the following model of growth and decay:

$$\frac{dx_i}{dt} = r_i x_i \left(1 - \frac{x_i}{Z_i}\right)^{\alpha_i} - q_i x_i^{\gamma_i} F_i. \tag{1}$$

$r_i$ is the natural growth rate of fish, $Z_i$ is the fishery's maximum carrying capacity, $F_i$ is the total amount of fishing in fishery $i$, and $q_i$ is the "catchability coefficient." When $\alpha_i = \gamma_i = 1$, this is the common Gordon-Schaeffer model for fisheries. $\alpha_i \neq 1$ allows the dependency between the natural growth rate and biomass to be nonlinear, and $\gamma_i \neq 1$ allows one to model "patchy" populations. Populations that are more patchy ($\gamma_i < 1$) remain relatively easy to catch when population declines because the remaining stocks stick together, while non-patchy populations become harder to catch as population declines because a smaller population is spread over the same geographic space. The long-term biomass, given $F_i$, is found by setting (1) equal to 0, which can be computationally solved virtually instantaneously. This long-term biomass will be denoted $\tilde{x}_i$.

Due to overlapping claims in the SCS, each player independently determines how many fishing quotas to issue for each fishery. These choices will be made with the aim of maximizing the sum of utilities from each fishery, which for Blue is defined as:

$$\pi_B(P_B, P_R) = \sum_{i=1:k} \pi_{B,i} \tag{2}$$

$$\pi_{B,i} = (p_i q_i \tilde{x}_i^{\gamma_i} - \psi_{B,i}) F_{B,i}, \tag{3}$$

where $P_B = [P_{B,1}, P_{B,2}, ..., P_{B,k}]$ and $P_R = [P_{R,1}, P_{R,2}, ..., P_{R,k}]$ hold Blue and Red's patrol allocations for each fishery, $F_{B,i}$ is the amount of quotas Blue allocated to fishery $i$, $p_i$ is the price fetched for one metric ton of fish from fishery $i$, and $\psi_{B,i}$ is the total cost of fishing for a Blue fishermen in fishery $i$.

$\psi_{B,i}$ of course accounts for standard operating and opportunity costs, but because the fisheries are in dispute costs imposed by patrols are also influential. The following linear model is used for costs:

$$\psi_{B,i} = c_B + \max\{0, \beta_{BR} P_{R,i} - \beta_{BB} P_{B,i}\}. \tag{4}$$

The term $\beta_{BR}$ quantifies the effect Red patrols impose on Blue fishermen, and $\beta_{BB}$ quantifies the ability of Blue patrols to offset that effect. The net effect of patrols on costs obviously cannot be negative, so the $\max\{\cdot\}$ function is used to keep costs at or above Blue's level of operating and opportunity costs, $c_B$. For simplicity it's been assumed a patrol allocated to fishery $i$ does not have the range to influence other fisheries. While in practice this may not be accurate, this assumption doesn't affect the methodological approach of Section 4. Equations (2), (3), and (4) can analogously be defined for Red.

Once the patrol allocations for each player are given, it's possible to derive Nash equilibrium fishing levels in each fishery if $\alpha_i = \gamma_i = 1$. In the general case, equilibrium levels can be modeled as a function of $(P_B, P_R)$ using a truncated polynomial model. To see this, consider the following:

**Result 1. Equilibrium fishing levels, given patrols**

1. Define $F^*_{B,i}(F_{R,i}, P_{B,i}, P_{R,i})$ and $F^*_{R,i}(F_{B,i}, P_{B,i}, P_{R,i})$ as the optimal fishing levels for Blue and Red, respectively, given the other's fishing levels and each's patrol allocation. Define $x_i(F_{B,i}, F_{R,i})$ as the biomass given fishing levels $F_{B,i}$ and $F_{R,i}$.

2. $F^*_{B,i}(F_{R,i}, P_{B,i}, P_{R,i}) = 0$ iff $p_i q_i x_i \left(0, F^*_{R,i}(0, P_{B,i}, P_{R,i})\right)^{\gamma_i} - \psi_{B,i} \leq 0$; that is, it's optimal for Blue to fish iff there is positive utility from her first unit of fishing. Otherwise, $F^*_{B,i}(F_{R,i}, P_{B,i}, P_{R,i})$ can be modeled using a non-computationally expensive design of experiment (DOE) over $(F_{R,i}, P_{B,i}, P_{R,i})$; a polynomial regression is seen to fit sampled data with accuracy $R^2 \approx .99$. The same holds for modeling $F^*_{R,i}(F_{B,i}, P_{B,i}, P_{R,i})$.

3. Given the near perfect models for optimal response fishing levels in point 2, a non-computationally expensive DOE can be performed over $(P_B, P_R)$ to find equilibrium fishing levels as a function of patrols. For each sample, first check if $F_{B,i}$ or $F_{R,i}$ are provably 0 (per step 1). If neither are, standard nonlinear methods can find an equilibrium in less than one second using the polynomial models for optimal responses; in this paper the L-BFGS-B method was used (Byrd et al. 1995). Fitting a further polynomial for equilibrium fishing levels as a function of patrols has accuracy $R^2 \approx .99$.

The sampling methods described in Result 1 don't add a material amount of time to the broader problem of optimally allocating patrol vessels across many fisheries, as described in Section 4. Denote the equilibrium fishing levels modeled via Result 1 as $\tilde{F}_{B,i}$ and $\tilde{F}_{R,i}$. Recalling that $\tilde{x}_i$ is a function of $F_i = F_{B,i} + F_{R,i}$, and $\tilde{F}_{B,i}$ a function of $P_{B,i}$ and $P_{R,i}$, Blue's utility (and analogously, Red's) can be written strictly in terms of patrols:

$$\pi_B(P_B, P_R) = \sum_i \left( p_i q_i^{\gamma_i} \tilde{x}_i - c_B - \max\{0,\ \beta_{BR} P_{R,i} - \beta_{BB} P_{B,i}\} \right) \tilde{F}_{B,i}. \tag{5}$$

*3.2 Optimal response function for patrols*

For a given Blue strategy, $P_B$, Red's optimal response function is:

$$P_R^*(P_B) := \underset{P_R}{\operatorname{argmax}}\, \pi_R(P_B, P_R)\ s.t. \sum_{i=1:k} P_{R,i} = P_R^{tot}, \tag{6}$$

where $P_R^{tot}$ is the total number of patrols at Red's disposal. An analogous function could be defined for Blue's optimal response to Red. This function is a costly optimization problem, where the cost stems from needing to compute $\tilde{x}$ computationally, the need to model equilibrium fishing levels via a high-degree polynomial, and the potentially high value of $k$. Specifically, when $k = 10$ this optimization require an average of over ten minutes to solve on a standard laptop, and when $k = 60$ can take over 2 hours. Whether the purpose of analysis is to seek a pure Nash equilibrium or analyze a sequential game (as will be done in Section 4), because no analytical solution to (6) exists it will need to be solved for several values of $P_B$ in order to analyze the game. An efficient sampling technique is paramount and is the subject of the next section.

# 4 Robust Formulation and Response Surface Methodology for the Fishing Dispute Game

*4.1 Sequential, robust game formulation*

For the remainder of this paper the perspective of Blue will be taken, and it will be assumed Blue is to make a decision on how to allocate her patrols knowing Red will observe her strategy and respond. It's further assumed Blue will not respond in turn, but rather stick to her original decision. Methodologically, this simplifies the problem by defining a clear stopping point for the game. In practice, while Blue maintains the ability to change strategies at any time the assumption made here is reasonable as nations prefer not to continuously update strategic decisions. Because the perspective of Blue is taken, her utility function is known, but there is behavioral uncertainty in Red's decision-making process for the reasons described in Section 1. Thus, the following robust game formulation will be used:

$$P_B^* = \underset{P_B}{\operatorname{argmax}}\, \pi_B(P_B, P_R') \tag{7}$$

$s.t.$

$$P_R^* = \underset{P_R}{\operatorname{argmax}}\, \pi_R(P_B, P_R) \ \ s.t. \ \ \sum_{i=1:k} P_{R,i} = P_R^{tot}$$

$$P_R' = \underset{P_R}{\operatorname{argmin}}\, \pi_B(P_B, P_R) \ \ s.t. \ \ \sum_{i=1:k} P_{R,i} = P_R^{tot} \ \text{and} \ \pi_R(P_B, P_R^*)/\pi_R(P_B, P_R') - 1 < \varepsilon$$

$$\sum_{i=1:k} P_{B,i} = P_B^{tot}$$

For any particular strategy of Blue, $P_B$, this formulation first finds the Red strategy that optimizes the assumed function for Red's utility; this model-defined optimal response is denoted $P_R^*$. Then, the formulation seeks the Red strategy that minimizes Blue's utility, subject to a constraint that Red's realized (model-defined) utility is within $\varepsilon$ of his optimal utility; this strategy is denoted $P_R'$, and $\varepsilon$ is a tuning parameter that could be set to, e.g. 5-20%. Blue's objective is to find the strategy, $P_B^*$, which maximize the utility she realizes when Red responds by playing $P_R'$. As a useful point of notation, the resultant utility $\pi_B(P_B, P_R')$ will be denoted $\bar{\bar{\pi}}_B(P_B)$, and will often be referred to as the "robust utility of $P_B$."

Qualitatively, (7) seeks Blue strategies which safeguard against worst-case Red responses while constraining Red's strategic options to those that seem reasonable. This is an attractive formulation when a decision maker is uncertain of her adversary's decision-making process, but believes she has a model that's reasonable and hence adversary strategies that produce markedly suboptimal model-defined utilities are unlikely. As will be seen in Section 5, where examples are presented, the impact of this robust formulation can be quite significant. Solutions to (7) will be benchmarked against a non-robust formulation, and it's not uncommon for realized utilities to be 17% greater in the robust formulation (refer to Figure 1).

*4.2. Response surface methodology for optimization (7)*

Optimization (7) has two constraints which are themselves optimizations, each of which has no analytical solution and is computationally expensive. A brute force search across the strategy space for $P_B$ is infeasible for decently sized problems (e.g. $k = 10$). In light of this burden, the following sampling method was developed which employs a response surface to iteratively select sample points for $P_B$:

**Algorithm 1**

1. Generate initial samples of $P_B$.

    a. Generate $n_{LHS}$ samples using a modified Latin hypercube sample. Specifically, generate points by allowing $P_{B,i} \in [0,1]$ for all fisheries $i$, and then scale each sample so $\sum_i P_{B,i} = P_B^{tot}$.

    b. Generate $n_{uni}$ samples based on unilateral responses computed via a fast heuristic. Specifically, define $\tilde{P}_B(P_R)$ and $\tilde{P}_R(P_B)$ as easily computable functions that give good responses for Blue to Red, and Red to Blue, respectively, though not necessarily optimal responses. Then: (i) generate a random sample of $P_B$; (ii) use the fast heuristics to allow Red and Blue to sequentially respond to one another four times; (iii) the sample point to be used is the resulting Blue strategy. This sampling approach could conceivably provide a more representative sample of strategies which are likely to be used.

    c. For each sample point from 1.a and 1.b, use a global optimizer to find $P_R^*$ and $P_R'$ as defined in (7), and hence compute $\bar{\bar{\pi}}_B(P_B)$.

2. Fit a response surface for $\bar{\bar{\pi}}_B(P_B)$ using currently drawn samples of $P_B$, denoted $\hat{\bar{\bar{\pi}}}_B(P_B)$. If the fit is adequate, go to step 3. Otherwise, generate one more sample using the method in 1.a, one more using the method of 1.b, and repeat step 2.

3. Generate the next sample, $P_B^{sample}$, using the response surface, $\hat{\bar{\bar{\pi}}}_B(P_B)$, as follows:

    a. Select a random fishery, $i$.

    b. Randomly select $\alpha \in [0,1]$, and randomly determine whether this will serve as an upper- or lower-bound on $P_{B,i}/P_B^{tot}$. WLOG, assume an upper-bound of $P_{B,i}/P_B^{tot} < \alpha$.

c. Solve the following optimization to select the next sample point:

$$P_B^{sample} = \underset{P_B}{\operatorname{argmax}} \hat{\bar{\bar{\pi}}}_B(P_B) \quad (8)$$

s.t.

$$P_{B,i}/P_B^{tot} < \alpha \quad (8a)$$

$$\sum_{i=1:k} P_{B,i} = P_B^{tot}. \quad (8b)$$

4. Compute the actual robust utility, $\bar{\bar{\pi}}_B(P_B^{sample})$, add this to the existing data, and refit $\hat{\bar{\bar{\pi}}}_B(P_B)$. Return to step 3 and continue until a predefined computational budget is exhausted.

5. Estimate the solution to (7) by solving $\hat{P}_B^* = \underset{P_B}{\operatorname{argmax}} \hat{\bar{\bar{\pi}}}_B(P_B)$.

When implementing this algorithm for several examples in Section 5, the following specifications were used. In steps 1c, 3c, and 5, the COBYLA optimization method was used with 20 random restarts, which is a derivative-free optimizer able to handle arbitrarily complex objective functions and constraints (Powell 1998). Similarly, in step 1b, COBYLA was used with no random restarts as a fast heuristic. A boosted regression tree was seen to fit well for $\hat{\bar{\bar{\pi}}}_B(P_B)$ (Hastie, Tibshirani, and Friedman 2009). The number of trees to used, their maximum depths, and the learning rate while boosting were chosen via cross validation, and the criterion determining whether the fit is adequate is a cross-validated $R^2$ exceeding 0.5. To save time, cross validation is only performed in step 2 (not in step 4 when recalibrating the surface). The total number of samples drawn for each example was capped at 300, and in step 1 values of $n_{LHS} = n_{uni} = 20$ were used.

The crucial step in Algorithm 1 is step 3, and deserves a bit of commentary. In this step, the next sample point is selected by optimizing the response surface for robust utility, $\hat{\bar{\bar{\pi}}}_B(P_B)$, the idea being that where $\hat{\bar{\bar{\pi}}}_B(P_B)$ is large, so is $\bar{\bar{\pi}}_B(P_B)$. However, there is imprecision in the response surface, and solving (8) without the artificially added bound, (8a), will likely cause inadequate exploration of the decision space, and risks settling at a suboptimal solution. For instance, assuming solving (8) without (8a) produces the sample point $P'_B$. Computing $\bar{\bar{\pi}}_B(P'_B)$ and recalibrating the response surface will slightly alter the surface, but likely not materially. Thus, solving (8) again with the updated response surface will produce a very similar answer to $P'_B$. This process will repeat, essentially limiting the sampling algorithm to $n_{LHS} + n_{uni} + 1$ draws. Unless $\hat{\bar{\bar{\pi}}}_B(P_B)$ had a very high $R^2$ after the first $n_{LHS} + n_{uni}$ samples, it's likely the algorithm is missing the global optimum. Introducing the artificial constraint (8a) enforces a dense exploration of the decision space, while still concentrating effort where robust utility is expected to be high.

## 5 Examples

### 5.1 Parameter values and benchmark solution

This section presents the results of 1000 examples with randomly assigned model parameter values. Subsection 5.2 details one specific instantiation of the parameters which motivates the need for robust analysis, and subsection 5.3 summarizes results across all examples. The 1000 instantiations of the parameters were determined as follows.

In all examples, $k = 10$ fisheries are assumed and $\varepsilon = 10\%$ in (7). The parameters $Z_i$, $r_i$, $q_i$, $\alpha_i$, and $\gamma_i$ are assumed homogenous across fisheries and simply denoted $Z$, $r$, $q$, $\alpha$, and $\gamma$. Values are randomly generated using independent uniform distributions; namely, $r \sim Uni(.3,.5)$, $q \sim Uni(.0001,.0003)$, $\alpha \sim Uni(.5,1.5)$, and $\gamma \sim Uni(.5,1.5)$. The price of fish in fishery 1 is $p_1 \sim Uni(1,2)$ billion USD, and the remaining prices are determined via $p_i = p_1 + \frac{i-1}{k-1} \cdot (\$3{,}000{,}000{,}000 - p_1)$ (i.e. linearly increasing prices from $p_1$ to \$3 billion). Operating plus opportunity costs, $c_B$ and $c_R$, are assumed equal and determined via $c_B = c_R \sim Uni(100000, 200000)$ USD. $Z = 1$ is fixed across all examples as a way of regularizing instantiations; that is, when $Z = 1$ and $k = 10$, instantiations of the other parameters generally lead to sensible results.

The remaining parameters to be specified are $P_B^{tot}$, $P_R^{tot}$, $\beta_{BR}$, $\beta_{RB}$, $\beta_{BB}$, and $\beta_{RR}$, and deserve a bit of commentary. Because part of the aim of this paper is to provide a tool for analyzing the SCS fishing dispute, with China (Red) in competition with SEAFDEC nations (Blue), $P_R^{tot} = 1000$ and $P_B^{tot} = 600$ are fixed across examples to reflect China's superiority in patrol craft. China has also heavily invested in maritime militias relative to the SEAFDEC nations, which should make their fishermen less vulnerable to rival patrols. Thus, the distributions $\beta_{BR} \sim Uni(150, 600)$ USD and $\beta_{RB} \sim Uni\left(\frac{\beta_{BR}}{2}, \beta_{BR}\right)$ USD are used to ensure $\beta_{RB} < \beta_{BR}$. Because it's not obvious the ability of Chinese patrols to mitigate the effect of SEAFDEC patrols is greater or less than that of SEAFDEC patrols to mitigate Chinese patrols, it's assumed $\beta_{BB} = \beta_{RR} = \frac{\beta_{BR} + \beta_{RB}}{4}$; the latter inequality implicitly assumes the mitigating effect of patrols is significantly less than the initial deterrent effect on fishermen.

Before presenting examples, it's necessary to provide an alternative solution concept to benchmark the robust formulation against (without this, it would be impossible to assess the quality of the solution because the results are strictly based on a sampling method with no reference to an analytical solution). The most natural benchmark is to simply assume Red responds to Blue with his model-defined optima. That is, the following is solved:

$$P_B^{NR} = \underset{P_B}{\mathrm{argmax}}\, \pi_B(P_B, P_R^*) \tag{9}$$

s.t.

$$P_R^* = \underset{P_R}{\mathrm{argmax}}\, \pi_R(P_B, P_R) \; s.t. \; \sum_{i=1:k} P_{R,i} = P_R^{tot}$$

$$\sum_{i=1:k} P_{B,i} = P_B^{tot}.$$

The superscript "NR" in $P_B^{NR}$ indicates this is a "non-robust" solution. As a point of notation, the resultant utility $\pi_B(P_B, P_R^*)$ will be denoted $\bar{\bar{\pi}}_B(P_B)$ and referred to as the non-robust utility of $P_B$. Like (7), this formulation has no analytical solution and is analyzed using a response surface methodology analogous to Algorithm 1. When comparing the robust solution, $P_B^*$, to the non-robust $P_B^{NR}$, two things should be noted. First, $\bar{\bar{\pi}}_B(P_B^*) - \bar{\bar{\pi}}_B(P_B^{NR})$ exceeding 0 on a consistent basis indicates Algorithm 1 is working properly. Second, if $\bar{\bar{\pi}}_B(P_B^*)$ is *significantly* greater than $\bar{\bar{\pi}}_B(P_B^{NR})$ for a particular instantiation of the parameters, this suggests behavioral uncertainty is a critical concern. As seen in Figure 1, $\bar{\bar{\pi}}_B(P_B^*) - \bar{\bar{\pi}}_B(P_B^{NR})$ consistently exceeds 0 and is occasionally quite large. The next subsection details an example where it's large.

*5.2 Example 1*

Before summarizing the results across all 1000 examples, the following is presented as a motivating example for the value of robust analysis. Consider the following instantiations of the parameters: $r = .36$, $q = .00015$, $\alpha = .91$, $\gamma = 1.06$, $p_1 = \$1,517,519,809.38$, $c_B = c_R = \$141,995.48$, $\beta_{BR} = \$579.56$, and $\beta_{RB} = 451.23$. Using Algorithm 1 to estimate $P_B^*$ and an analogous algorithm for $P_B^{NR}$ produces the results in Table 1.

**Table 1**
**Solutions Found via Response Surface Algorithms for Example 1**

| Fishery | 1 | 2 | 3 | 4 | 5 | 6 | 7 | 8 | 9 | 10 |
|---|---|---|---|---|---|---|---|---|---|---|
| $P_B^*$ | 19.31 | 22.61 | 16.53 | 8.82 | 56.55 | 48.60 | 47.62 | 15.26 | 70.02 | 294.68 |
| $P_B^{NR}$ | 16.74 | 46.57 | 6.45 | 0.00 | 43.49 | 65.93 | 49.93 | 99.82 | 127.71 | 143.37 |

The robust utilities of these two solutions are $\bar{\bar{\pi}}_B(P_B^*) = \$166,663,708.36$ and $\bar{\bar{\pi}}_B(P_B^{NR}) = \$130,857,172.92$. The robust utility of $P_B^*$ is 27.36% higher than that of $P_B^{NR}$, clearly indicating the downside risk Blue faces of assuming Red responds perfectly in accord with the assumed model for his behavior. In contrast, the non-robust utilities of these solutions (that it, those occurring when Red does play his model-defined optimal response), are $\bar{\pi}_B(P_B^*) = \$196,247,780.94$ and $\bar{\pi}_B(P_B^{NR}) = \$197,344,455.69$, a difference of less than 1%. Robust optimization in general can be fairly criticized as being too risk averse by focusing on worst-case results; in this particular example, however, there's little downside risk in playing $P_B^*$.

Figures 1 and 2 show that the general finding seen here, where the robust patrol allocation $P_B^*$ offers significant protection against behavioral uncertainty without diminishing the realized utility if Red plays his model-defined optimum, are not typical, but not terribly unusual either. The purpose of this example was to illustrate how important behavioral uncertainty and the robust formulation *can* be; Section 5.3 comments on how a decision maker may assess whether the robust formulation (7), or the non-robust (9) is the more appropriate decision-making tool.

*5.3 Additional examples*

The results of all 1000 examples are summarized in Figures 1 and 2. As in Example 1, the importance of accounting for behavioral uncertainty is assessed by comparing $\bar{\bar{\pi}}_B(P_B^*)$ to $\bar{\bar{\pi}}_B(P_B^{NR})$. This comparison is summarized in the metric $v := \bar{\bar{\pi}}_B(P_B^*)/\bar{\bar{\pi}}_B(P_B^{NR}) - 1$. Also in line with Example 1, the potential downside of using the robust solution is assessed. This is captured in the metric $w := v + (\bar{\pi}_B(P_B^*)/\bar{\pi}_B(P_B^{NR}) - 1)$, which correctly penalizes cases where significant utility is foregone by playing $P_B^*$ instead of $P_B^{NR}$, provided Red responds by playing his model-defined optimum.

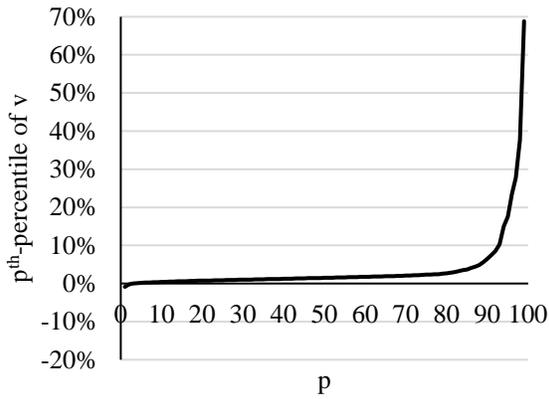

Fig. 1. Percentiles of $v$

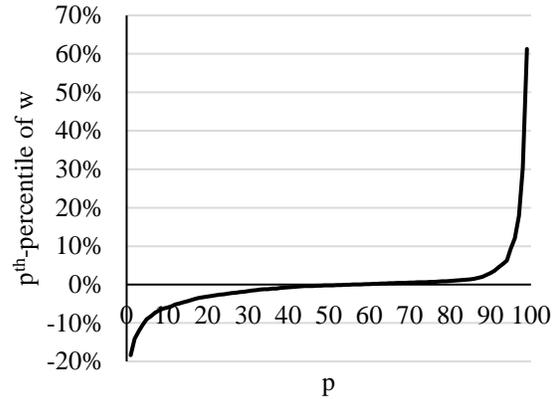

Fig. 2. Percentiles of $w$

As these figures show, $v$ and $w$ can become quite large. The 95th percentiles, for example, are 17.52% and 9.62%, respectively, indicating a robust solution to behavioral uncertainty may be preferable even after considering the possibility the non-robust solution offers better utility when Red responds per the model. In practice, a decision maker could use the response surface algorithm developed in this paper to estimate both $P_B^*$ and $P_B^{NR}$, compute $v$ and $w$, and use these results to subjectively determine whether to play $P_B^*$ and $P_B^{NR}$. A more rigorous approach would be to optimize a convex combination of robust and non-robust utilities in Algorithm 1, namely, $\lambda \cdot \pi_B(P_B, P_R') + (1 - \lambda) \cdot \pi_B(P_B, P_R^*)$, where $\lambda \in [0,1]$ is chosen based on how heavily the decision maker values robustness. This formulation won't add a material amount of computation time as $P_R'$ and $P_R^*$ were already required for each sampled $P_B$ when using the algorithm to optimize robust utility.

## 6 Conclusion and Future Research

This paper introduced a complex model for the fishing dispute in the South China Sea, where the strategic allocation of maritime patrols for two players impacts realized utilities. The analysis took a decision-theoretic approach where one player, Blue, needed to make a decision while accounting for uncertainty in Red's response. Rather than using a stochastic model with clearly specified probability distributions governing Red, a robust model was developed which requires only one parameter, $\varepsilon$, to specify how far Red is allowed to deviate from a model-defined optimum.

The formulated game is complex in the sense that an analytical solution doesn't exist and can only be analyzed by drawing computationally expensive samples. In addition to the above-mentioned robust formulation, the computational difficulty is driven by a highly nonlinear fisheries model, a budget constraint on patrols, and the presence of numerous fisheries in the SCS. A sampling algorithm incorporating RSM was developed to successfully estimate solutions to the game. The algorithm is quite flexible as it uses a derivative-free approach to select sample points, and can thus serve as an estimation technique for other arbitrarily complex games.

This paper didn't attempt to derive the traditional theoretical results found in the RSM literature, such as guarantees for convergence, based on the presumption that sufficiently complex problems are better modeled by complex response surfaces that won't lend themselves to such analytical results. In this case, a boosted regression tree was used but similarly complex surfaces are available. A strand of future research would be to perform extensive experiments on a wide variety of problems, possessing a variety of overarching features, and attempt to identify factors where certain types of response surfaces perform well. For example, a tree-based method performed well in this game of strategic resource allocation. This may persist into other games of strategic resource allocation as optimal behavior may suddenly shift in a discontinuous way once one battlefield becomes dominated by one player.

One other point of future research is to operationalize the methodology developed here by engaging fisheries experts to truly model the SCS. This paper was experimental and tested an RSM-based methodology across several instantiations of fishery parameters. It also assumed fisheries were dispersed enough such that each patrol vessel could be allocated to one and only one fishery at a time. A virtue of the methodology used in this paper is no assumptions were imposed on the underlying model, hence this latter point requires no modification to the methodology. Another empirical point, mentioned earlier, is that research must still be performed to understand how patrols impose costs on fishermen in a disputed fishery; currently, the best research has only examined the effect of patrols on illegal fishermen under undisputed territorial rights. By accurately modeling the parameters governing the SCS, and the effects of patrols on costs, policy makers will have a valuable tool for answer critical questions surrounding this fisheries dispute.

**Statements and Declarations**

*Funding and Competing Interests*

No funding was received to assist with the preparation of this manuscript. The author has no conflicts of interest to declare that are relevant to the content of this article.

*Data Availability*

The data generated for this study are available from the corresponding author upon request.